\documentstyle[preprint,aps]{revtex}

\begin{document}
\title{Periodic Orbit Quantization beyond Semiclassics}
\author{G\'abor Vattay\cite{LAbs}}
\address{Division de Physique Th\'eorique\cite{PU}, Institut de
Physique Nucl\'eaire,\\
F-91406 Orsay Cedex, France\\
}
\author{Per E. Rosenqvist}
\address{Center for Chaos and Turbulence Studies,
	 Niels Bohr Institute,\\
Blegdamsvej 17, DK-2100  Copenhagen \O, Denmark\\
}
\date{\today}

\maketitle

\begin{abstract}
A quantum generalization of the semiclassical theory of
Gutzwiller is given. The new formulation leads to
systematic orbit-by-orbit inclusion of higher $\hbar$ contributions
to the spectral determinant. We apply the theory to billiard systems,
and compare the periodic orbit quantization including the first $\hbar$
contribution to the exact quantum mechanical results.
\end{abstract}

The Gutzwiller trace formula\cite{Gutzwiller} is the most compact formulation
of the
semiclassical quantization of classically chaotic systems.
It  expresses the trace of the
Green function in terms of classical  periodic orbits of the system. In recent
years
it has been demonstrated\cite{Gutzwiller,Chaos} on many classically chaotic
systems
that this is indeed a very good approximation. Nevertheless, it is only the
leading order
formula in $\hbar$ and systematic inclusion of higher order $\hbar$
contributions is desirable.
In this letter we present a new method, which simplifies the computation
of these contributions and gives some insight in their structure.
Our key result is the generalized Selberg product (\ref{SPD}) which expresses
the global eigenvalues in terms of local eigenspectra computed on neighborhoods
of individual periodic orbits.

The trace of the energy domain Green function is
\begin{equation}
\mbox{Tr}\, G(x,x',E) =\int dx \, G(x,x,E)=\sum_n \frac{1}{E-E_n},\label{tgf}
\end{equation}
where $E_n$ are the eigenenergies of the system.  The trace is a logarithmic
derivative of the spectral determinant
\begin{equation}
\mbox{Tr}\,G(x,x',E)=\frac{d}{dE} \log \Delta(E),\label{2}
\end{equation}
where $\Delta(E)=\det(E-\hat{H})=\prod_n (E-E_n)$
is the formal secular determinant of the Hamilton operator $\hat{H}$.
The Green function\cite{Rc} can be written\cite{Balian1} as a sum over
classical trajectories connecting $x$ and $x'$ with energy $E$
$$
G(x,x',E)=\sum_{cl}A(x,x',E)e^{iS(x,x',E)/\hbar},
$$
where $S(x,x',E)$ is the classical action of the trajectory, the amplitude
$A(x,x',E)$ can be determined by the multiple scattering
expansion.
For $\hbar\rightarrow 0$ the trace of the Green function can
be evaluated with the saddle point method. The saddles of the function
$S(x,x,E)$ are the classical primitive periodic orbits $p$
with energy $E$, and their repetitions $r$.
Consider the trace of the Green function restricted to
a small tube $\Omega_p$ enclosing the periodic orbit $p$:
$$\mbox{Tr}\, G_p(E)=
\int_{\Omega_p} d\tau dq_1dq_2...dq_{d-1} G(\tau,{\bf q},\tau,{\bf q},E),
$$
where $\tau$ is a local coordinate along the orbit and $q_i,i=1,...,d-1$ are
coordinates orthogonal to the orbit.
We shall expand the action $S_{p,r}(\tau,{\bf q},\tau',{\bf q}')$ and
the amplitude $A_{p,r}(\tau,{\bf q},\tau',{\bf q}')$ around the periodic orbit
and repetitions
in Taylor series and $\mbox{Tr}\, G_p(E)$ can be written as a trace of the
locally
expanded Green's function
$G_p(\tau,{\bf q},\tau',{\bf q}',E)=\sum_{p,r}
A_{p,r}(\tau,{\bf q},\tau',{\bf q}',E)
e^{iS_{p,r}(\tau,{\bf q},\tau',{\bf q}',E)/\hbar}.
$
The trace of the full Green function can be expressed with these
local traces
\begin{equation}
\mbox{Tr}\, G(E)=\mbox{Tr}\, G_0(E) + \sum_p \mbox{Tr}\, G_p(E),\label{5}
\end{equation}
where
$Tr G_0(E)$ is the Weyl or Thomas-Fermi term\cite{Balian2} coming
from zero length trajectories.
For $\hbar\rightarrow 0$ the values of the integrals are independent
of the sizes and overlaps of the $\Omega_p$ tubes. As we shall be
expanding around the classical orbits, our results are valid only in
the sense of asymptotic series;  finite $\hbar$ computations
require care, especially for non-hyperbolic or
non-elliptic situations where periodic orbits might accumulate.
In applications the number of periodic orbits used should depend on
$\hbar$ and on the order of truncation of the Taylor expansion of the action
and amplitude. Berry and Keating\cite{beke} have proposed a method
for dealing with this problem; here we assume that our computations
(here only the first $\hbar$ contribution will be evaluated) are
within the scope of the validity of saddle point expansions.

The standard way of evaluating the above integrals is the Feynman
diagram perturbative exapansion about the saddle point, implemented
in ref.~\cite{Gaspard}. However, this is not the only way to
compute the trace of $G_p$; here we offer an alternative,
based on computing the local eigenspectrum. Since $G_p$ is the locally expanded
form
of the Green function of the Schr\"odinger equation, we can assume that
there exist a suitable local expansion of the Schr\"odinger equation
whose Green function is $G_p$.
If we can solve the local Schr\"odinger problem on $\Omega_p$,
then the trace of $G_p$  can be recovered as a logarithmic derivative of the
spectral determinant $\Delta_p(E)$ of the local eigenvalue problem
\begin{equation}
\mbox{Tr}\, G_p(E)=\frac{d}{dE} \log \Delta_p(E).\label{6}
\end{equation}
To define the local Schr\"odinger problem we introduce a
representation of the Schr\"odinger equation which is closely related to
classical dynamics and facilitates development of a semiclassical expansion
around a given classical orbit\cite{Sch}. We take an arbitrary solution of the
Hamilton-Jacobi equation
\begin{equation}
\partial_t S(x,t) +\frac{1}{2}(\nabla S(x,t))^2 + U(x)=0,\label{hamjac}
\end{equation}
and transform the wave function $\psi(x,t)$ of the Schr\"odinger
equation to $\varphi(x,t)=\psi(x,t)e^{-iS(x,t)/\hbar}$. The new wave function
satisfies
the Maslov-Fjedoriuk transport equation\cite{MasFjed}
\begin{equation}
\partial_t\varphi+\nabla(\varphi\nabla S)
-\frac{1}{2}\varphi\Delta S
-\frac{i\hbar}{2}\Delta\varphi=0\label{ampl}.
\end{equation}
The quantum mechanical propagator in this representation is
\begin{equation}
\langle x',t'|x,t \rangle =e^{\frac{i}{\hbar}(S(x,t)-S(x',t'))}
{\cal L}(x',t'|x,t),
\label{propa}
\end{equation}
where ${\cal L}(x',t'|x,t)$ is the propagator of the transport equation
(\ref{ampl}).
By the analogy between (\ref{ampl}) and the Fokker-Planck equation we can write
this propagator as a sum over stochastic trajectories fluctuating around
classical
orbits
$$
{\cal L}(x',t'|x,t)=\left\langle e^{\frac{1}{2}\int_t^{t'}\Delta
S(\xi(t'',t,x),t'')dt''}\delta(x'-\xi(t',t,x))\right\rangle_{\eta}
\label{L}
$$
where $\xi(t',t,x)$ is the solution of the complex Langevin equation
\begin{equation}
\frac{d\xi}{dt'}=\nabla S(\xi,t') + \sqrt{i\hbar}\eta(t')\label{Langevin}
\end{equation}
with initial condition $\xi(t)=x$, and $\eta(t')$ is Gaussian white noise
with $ \langle \eta_i(t)\rangle =0$ and
$\langle \eta_i(t)\eta_j(t')\rangle=\delta_{ij}\delta(t-t')$
{}.
The symbol $\left\langle \right\rangle_{\eta}$ denotes averaging over all noise
configurations.
This representation based on an arbitrary solution of
the Hamilton-Jacobi equation has been
proposed in ref.~\cite{Ron} as an alternative to the Feynman path
sum\cite{Schulman}.
As $\hbar\rightarrow 0$ the fluctuations are suppressed,
(\ref{Langevin}) becomes a deterministic
equation and $\xi(t',t,x)$ becomes the classical trajectory with initial
coordinate
$x$ and momentum $p(t)=\nabla S(x,t)$.
We develop a multiple scattering expansion of the propagator
by handling the diffusive term $-\frac{i\hbar}{2}\Delta\varphi$ in (\ref{ampl})
perturbatively
$$
{\cal L}(x',t'|x,t)={\cal L}^{(0)}(x',t'|x,t)+
\sum_{n=1}^{\infty}(i\hbar/2)^n {\cal L}^{(n)}(x',t'|x,t),
$$
where
$$
{\cal L}^{(0)}(x',t'|x,t)=e^{\frac{1}{2}
\int_t^{t'}\Delta S(\xi(t'',t,x),t'')dt''}\delta(x'-\xi(t',t,x))
$$
and $\xi(t',t,x)$ is the deterministic trajectory.
To study the propagator (\ref{propa}) for  $x,x'\in\Omega_p$ we can choose
$S_p(x,t)$ such
that the momentum $p(t)=\nabla S_p(x_p(t),t)$ coincides with the momentum of
the periodic orbit $p_p(t)$ and the velocity field $\nabla S_p(x,t)$
is independent from $t$. Such a local velocity field can be
constructed order by order by substituting the the Taylor expansion
\begin{equation}
S_p(x,t)=\sum_{\bf n}^{\infty} s_{\bf n}(t)
(x-x_p(t))^{\bf n}/{\bf n!},\label{hatas}
\end{equation}
into the Hamilton-Jacobi equation (\ref{hamjac}).
The symbol ${\bf n}=(n_1,n_2,...,n_d)$ denotes the multiple index
in $d$ dimensions, ${\bf n!}=\prod_{i=1}^{d} n_i!$,
and $(x-x_p(t))^{{\bf n}}=\prod_{i=1}^{d}(x_i-x_{p,i}(t))^{n_i}$.
Due to the time independence of the velocity field,
the expansion coefficients are periodic functions $s_{\bf n}(t)
=s_{\bf n}(t+T_p)$,
except the zero order term which grows with each period by
$s_0(t+T_p)-s_0(t)=\int_0^{T_p} \left(\frac{1}{2}p_p^2(t')-U(x_p(t'))
\right)dt'
=\int_0^{T_p}L(t')dt'$. This solution is stationary $\partial_t S_p(x,t)=-E$
due to the time independent velocity field. By order-by-order
construction in each step we get
a formally periodic  velocity field, therefore it is advantageous to
study  the eigenvalue problem
using the one-period operator $\langle x',T_p|x,0 \rangle_p$, where $p$
indicates the computation of the propagator with the above procedure
using $S_p(x,t)$.
If $\psi_p(x,t)$ is a stationary state of the local problem, then it is also
an eigenfunction of the one-period propagator
\begin{equation}
e^{-iET_p/\hbar}\psi_p(x,0)=\int dx'\langle x,T_p|x',0\rangle_p \psi_p(x',0).
\end{equation}
Using the fact that $S_p(x,T_p)-S_p(x,0)=\int_0^{T_p} dt'L(t')$ and that the
action of the periodic orbit is
$S_p(E)=\oint {\bf p}_p {\cdot} d{\bf q}=\int_0^{T_p} L(t')dt'+ET_p$
we get for the transformed wave function
\begin{equation}
\varphi_p(x,0)=e^{iS_p(E)/\hbar}\int dx'{\cal L}_p(x,T_p|x',0)\varphi_p(x',0).
\end{equation}
If the eigenvalues $e^{\lambda_{p,{\bf m}}(E)}$ of ${\cal L}_p$ are known
\begin{equation}
e^{\lambda_{p,{\bf m}}(E)}\varphi_{p,{\bf m}}(x,0)=\int dx' {\cal
L}_p(x,T_p|x',0)
\varphi_{p,{\bf m}}(x',0),\label{loceig3}
\end{equation}
where ${\bf m}$ is the multiple index labeling the eigenstate,
the  local spectral determinant is given by
\begin{equation}
\Delta_p(E)=\prod_{\bf m}(1-e^{\lambda_{p,{\bf m}}(E)}
e^{\frac{i}{\hbar}S_p(E)}).
\end{equation}

Now we can reexpress the spectral determinant of the full problem
through (\ref{2}),(\ref{5}) and (\ref{6})
\begin{equation}
\Delta(E)=e^{W(E)}\prod_p\prod_{\bf m}(1-e^{\lambda_{p,{\bf m}}(E)}
e^{\frac{i}{\hbar}S_p(E)})
\label{SPD},
\end{equation}
where $W(E)=\int^E dE' \, \mbox{Tr}\,G_0(E')$ is the Weyl term contribution.
This Selberg-product type expression for the spectral determinant
is our main result; it is the {full quantum} generalization of the
semiclassical Gutzwiller-Voros product formula\cite{Voros}. A similar
decomposition has been found for quantum baker maps in ref.~\cite{Saraceno}.
The functions
\begin{equation}
\zeta^{-1}_{\bf m}(E)=\prod_p \left(1-e^{\lambda_{p,{\bf m}}(E)}
	e^{\frac{i}{\hbar}S_p(E)}\right)\label{Ruelle}
\end{equation}
are the generalizations of the Ruelle dynamical
zeta functions \cite{Ruelle}.
The corresponding generalization of the
semiclassical Gutzwiller trace formula to the full quantum
mechanical trace follows from (\ref{2}); it
can be writen as a sum over the prime cycle repetition index $r$:
\begin{equation}
\mbox{Tr}\, G(E)=\mbox{Tr}\, G_0(E)+\frac{1}{i\hbar}\sum_{p,r,{\bf
m}}\left[T_p(E)-i\hbar\frac{d  \lambda_{p,{\bf m}}(E)}{dE}\right]
\left(e^{\lambda_{p,{\bf m}}(E)} e^{\frac{i}{\hbar}S_p(E)}\right)^r.
\end{equation}

In practice the eigenvalues of (\ref{loceig3}) can be
computed the following way. We substitute the perturbation
expansion
$
\varphi_{p,{\bf m}}(x,t)=\sum_{\ell=0}^{\infty}
\left(\frac{i\hbar}{2}\right)^\ell
\varphi_{p,{\bf m}}^{(\ell)}(x,t)
$
in the equation (\ref{ampl}) and
we get an
iterative scheme starting with the semiclassical
solution $\varphi_{p,{\bf m}}^{(0)}$:
\begin{eqnarray}
\partial_t\varphi_{p,{\bf m}}^{(0)}+\nabla\varphi_{p,{\bf m}}^{(0)}\nabla S_p+
\frac{1}{2}\varphi_{p,{\bf m}}^{(0)}\Delta S_p&=&0,\\ \nonumber
\partial_t\varphi_{p,{\bf m}}^{(\ell+1)}+\nabla\varphi_{p,{\bf
m}}^{(\ell+1)}\nabla S_p+
\frac{1}{2}\varphi_{p,{\bf m}}^{(\ell+1)}\Delta
S_p&=&\Delta\varphi_{p,{\bf m}}^{(\ell)}.
\label{coupeq}
\end{eqnarray}
The eigenvalue can also be expanded in powers of
$i\hbar/2$:
\begin{eqnarray}
\lambda_{p,{\bf m}}(E)&=&\sum_{\ell=0}^{\infty}
\left(\frac{i\hbar}{2}\right)^\ell C_{p,{\bf m}}^{(\ell)}\\
\nonumber .
\end{eqnarray}
The eigenvalue equation (\ref{loceig3}) in $\hbar$
expanded form reads as follows
\begin{eqnarray}
\varphi_{p,{\bf m}}^{(0)}(x,T_p)&=&\exp(C_{p,{\bf m}}^{(0)})\varphi_{p,{\bf
m}}^{(0)}(x,0), \\
\varphi_{p,{\bf m}}^{(1)}(x,T_p)&=&\exp(C_{p,{\bf
m}}^{(0)})\left\{\varphi_{p,{\bf m}}^{(1)}(x,0)+C_{p,{\bf m}}^{(1)}
\varphi_{p,{\bf m}}^{(0)}(x,0)\right\},
\nonumber
\end{eqnarray} and so on \cite{long}.
This approach is very similar to the one used in ref.~\cite{Babic}.

We expand the functions $\varphi_{p,{\bf m}}^{(\ell)}(x,t)$
as in (\ref{hatas}), in Taylor series around the periodic orbit, and to solve
recursively the equations (\ref{coupeq}). Only a couple of
coefficients should be computed to derive the first
corrections; the detailed calcualtion will be published
elsewhere\cite{long}.
The zeroth order term yields the Gutzwiller semiclassical weight
$C_{p,{\bf m}}^{(0)}=-i\pi\nu_p
-\sum_{i=1}^{d-1}\left(m_i+\frac{1}{2}\right)u_{p,i},$
where $u_{p,i}=\log |\Lambda_{p,i}|$ are the
Lyapunov exponents of $p$, $|\Lambda_{p,i}$ the eigenvalues
of the monodromy matrix ${\bf J}_p$, and $\nu_p$ is the Maslov index
of the periodic orbit. The first correction is given by the integral
$C_{p,{\bf m}}^{(1)}=\int_0^{T_p}dt\, \Delta
\varphi_{p,{\bf m}}^{(0)}(t)/\varphi_{p,{\bf m}}^{(0)}(t).$

The semiclassical trace formula can be recovered by
dropping the subleading
term $-i\hbar {d \over dE}\lambda_{p,{\bf m}}(E)$ and using the
semiclassical
eigenvalue
$e^{\lambda^{(0)}_{p,\bf m}(E)}=e^{C_{p,{\bf m}}^{(0)}}=e^{-i\nu_p\pi -\sum_i
(m_i+1/2)u_{p,i}}$.
Summation for the indexes $m_i$ yields the
semiclassical amplitude
\begin{equation}
\sum_{{\bf m}}e^{\lambda^{(0)}_{p,{\bf m}}(E) r}=\frac{e^{-ir\nu_p\pi}}{\mid
\det({\bf
1}-{\bf J}_p^r)\mid^{1/2}}.
\label{gutzy}
\end{equation}

When the theory is applied to billiard systems,
the wave function should fulfill the Dirichlet boundary
condition on hard walls.
The wave function determined from (\ref{ampl})
behaves discontinuously when the trajectory $x_p(t)$
hits the
wall. For simplicity we consider a two dimensional
billiard system
here.
The wave function on the wall at the time
$t_{-0}$ immediately before the bounce is given by
\begin{equation}
\psi_{in}(x,y(x),t)=\varphi(x,y(x),t_{-0})e^{iS(x,y(x),t
_{-0})/\hbar},
\end{equation}
where $y(x)=Y_2x^2/2!+Y_3x^3/3!+Y_4x^4/4!+...$ is the
parameterization of
the wall around the point of reflection.
The wave function on the wall at the time
$t_{+0}$ immediately after the bounce is given by
\begin{equation}
\psi_{out}(x,y(x),t)=\varphi(x,y(x),t_{+0})
e^{iS(x,y(x),t_{+0})/\hbar}.
\end{equation}
The sum of these wave functions should vanish
on the hard wall.
This implies that the incoming and the outgoing
amplitudes and the phases are related as
$S(x,y(x),t_{-0})=S(x,y(x),t_{+0})$ and
$\varphi(x,y(x),t_{-0})=-\varphi(x,y(x),t_{+0})$.
The minus sign can be interpreted as the Maslov phase
associated with the hard wall reflection.

In order to gauge the improvement due to the inclusion  of
the quantum corrections we have developed
a numerical code\cite{Program} for calculating the
first correction $C_{p,{\bf m}}^{(1)}$ for general
two-dimensional billiard systems. The first correction
depends only on the basic periodic orbit data
such as the lengths of the free flights
between bounces, the incidence angles and
the first three Taylor expansion coefficients
$Y_2,Y_3,Y_4$ of the wall in the point of incidence.
To check that our new local method gives
the same result as the direct calculation of the Feynman
integral, we computed the first $\hbar$ correction
$C_{p,0}^{(1)}$ for the periodic orbits of the three
disk scattering system
\cite{Rice,CE}.
We have found agreement to the five decimal digits
listed for the quantum corrections calculated in ref.~\cite{Gaspard}
(our method generates these numbers to any desired precision).
The $m\neq 0$ coefficients cannot be
compared to ref.~\cite{Gaspard}, since the $m$ dependence
was not realized there due to the lack of general
formulas (\ref{SPD}) and (\ref{Ruelle}). However, the
$m$ dependence can be
checked on the 2 disk
scattering system\cite{Wirzba}. For the standard
example\cite{Rice,CE,Gaspard,Wirzba}, with the distance
of the centers ($R$) taken to be 6 times the disk radius ($a$),
we obtain
$C_{p,{\bf m}}^{(1)}=(  -0.625 m^3 - 0.3125 m^2 + 1.4375 m +  0.625
)/\sqrt{2E}.$
For $m=0$ and $1$ this has been confirmed by A.
Wirzba\cite{AW}, who was able to extract these terms
from his exact quantum calculation.
Furthermore, our calculation has been verified for the
confocal hyperbola system at many different parameters\cite{WH}.
Our method makes it possible to utilize the symmetry
reduction and to include quantum coreections
into the fundamental domain cycle expansion
calculation of ref.~\cite{CE}.
We have computed the first correction to the leading 226 primitive
periodic orbits with 10 or less bounces in the
fundamental domain.
Table I demonstrates that the error of the corrected calculation
vs. the error of the semiclassical calculation
decreases with the wave number. Besides increasing the accuracy, a
fast convergence up to six decimal digits can be observed,
in contrast to the three decimal digits obtained in the full domain
calculation\cite{Gaspard}.

The authors are grateful to E. Bogomolny, B. Eckhardt, P. Gaspard,
M. Sieber, P. Sz\'epfalusy, N. Whelan and A. Wirzba for discussions and
comments on the original manuscript. We thank P. Cvitanovi\'c the
conceptual reedition of our manuscript. G. V. thanks EU PECO for fellowship
during his stay in Orsay. P. E. R. was supported by SNF. This work was
partially supported by OTKA F 17166 and T 17493, and the Foundation for
the Hungarian Higher Education.

\begin{table}
\caption{Real part of the resonances (Re $k$)
of the three disk scattering system at disk separation
6:1. Semiclassical and first corrected cycle expansion
versus exact quantum calculation and the
semiclassical error $\delta_{SC} = |$Re$(k_{QM})- $ Re$(k_{SC})|$
divided by the error of
the first correction $\delta_{Corr}$. The magnitude of
the error in the imaginary  part of the resonances remains
unchanged.}
\begin{tabular}{| c|c|c|c|}
Quantum  &   Semiclassical   &  First correction   &
$\delta_{SC}/\delta_{Corr}$ \\ \hline
0.697995 &  0.758313   &   0.585150  &    0.53   \\
 2.239601 &  2.274278   &   2.222930  &    2.08   \\
 3.762686 &  3.787876   &   3.756594  &    4.13   \\
 5.275666 &  5.296067   &   5.272627  &    6.71   \\
 6.776066 &  6.793636   &   6.774061  &    8.76   \\
 ...     &  ...        &    ...      &    ...    \\
 30.24130 &  30.24555   &   30.24125  &    92.3   \\
 31.72739 &  31.73148   &   31.72734  &    83.8   \\
 32.30110 &  32.30391   &   32.30095  &    20.0   \\
 33.21053 &  33.21446   &   33.21048  &    79.4   \\
 33.85222 &  33.85493   &   33.85211  &    25.2   \\
 34.69157 &  34.69534   &   34.69152  &    77.0   \\
\end{tabular}
\end{table}

\end{document}